\documentclass[12pt]{article}

\usepackage{graphicx}
\usepackage{bm}
\usepackage{amsmath}
\usepackage{amssymb}

\begin{document}

\title{Theoretical model for the evolution of the linguistic diversity}
\author{Viviane M. de Oliveira\thanks{viviane@df.ufpe.br}, M. A. F. Gomes\thanks{mafg@ufpe.br} and I. R. Tsang\thanks{inden@ufpe.br}}

\maketitle

\noindent
Departamento de F\'{\i}sica, Universidade Federal de Pernambuco, 50670-901, Recife, PE, Brazil

\begin{abstract}
Here we describe how some important scaling laws observed in the distribution of languages on Earth can emerge from a simple computer simulation. The proposed language dynamics includes processes of selective geographic colonization, linguistic anomalous diffusion and mutation, and interaction among populations that occupy different regions. It is found that the dependence of the linguistic diversity on the area after colonization displays two power law regimes, both described by critical exponents which are dependent on the mutation probability. Most importantly for the future prospect of world's population, our results show that the linguistic diversity always decrease to an asymptotic very small value if large areas and sufficiently long times of interaction among populations are considered.
\end{abstract}

\section{Introduction}

The origins of the languages have been an issue of investigation and broad interest since ancient times, and recent advances in archeology, genetics and linguistics have been important to a better comprehension of the linguistic diversification \cite{Renfrew2}. However, there is not a universal consensus concerning the evolution of this diversity  \cite{Renfrew}. Some similarities among distinct groups of languages suggest that they must have a common ancestor. By comparing languages that belong to a same family, linguists try to construct the hypothetical ancestor language. According to the {\it out of Africa hypothesis}, the modern human beings originated in Africa about 100,000 years ago and substituted all the populations outside Africa. This  hypothesis receives strong confirmation from the family tree based on a sampling of nuclear DNA from a number of living populations \cite{Cavalli-Sforza2, Renfrew3}. Molecular genetics can also give us some insights regarding to the distribution of languages on the Earth. Cavalli-Sforza \cite{Cavalli-Sforza} compared the family tree which one obtains from molecular genetic data at a world level with a family tree established using only linguistic data and his results indicate a fair degree of overlap.

The future of languages is a matter of interest and concern as well \cite{Diamond}. It is estimated that at least $50\%$ of the existing languages may be extinct in the next century \cite{Sampat, Graddol}. While one hundred of languages are spoken by about $90\%$ of the world population, most languages are present in a single or in a few small regions. The loss of linguistic diversity is a subject of worry not only by the linguists, because languages provide an important way to better understand the past of our species. Even more, since some languages possess a very elaborated vocabulary to describe the world, the loss would imply also the loss of ecological knowledge.

By analyzing all the approximately 6,700 languages on Earth, Gomes et al \cite{Gomes1} showed that (i) the language diversity scales with area according to a power law $D \sim A^z$, where $z=0.41 \pm 0.03$, over almost six decades, and (ii) the number of languages $n$ spoken by a population of size larger than $N$, $n(>N)$, also display a power law behaviour: $n(>N) \sim N^{-\tau}$. The critical exponent $z$ is comparable to the ones we observe in ecology for the relationships between species diversity and area, which are usually in the range 0.1 to 0.45 \cite{Rosenzweig, Pelletier, Hanski}.

Here we study the evolution of the linguistic diversity by introducing a spatial model which considers the underlying diffusion mechanisms that generate and sustain this diversity. The model is used to describe the occupation of a given area by populations speaking various languages. In the process of colonization of regions, language mutation or differentiation and language substitution can take place, and so increase the linguistic diversity. In the context of language dynamics, mutations are variations of languages with respect to a common ancestor language. The probability of producing reverse mutations is zero, that is, the language generated by a mutation is always different of the previous ones.

\section{Model}

Our model is defined on a two-dimensional lattice composed by $A=L \times L$ sites with periodic boundary conditions. Each lattice site $s_i$ represents a region that can be occupied by a population speaking just one language. We ascribe to each site a given capability $C_i$, whose value we estimate from a uniform distribution, in the range 0-1. This capability means the amount of resources available to the population which will colonize that place. It is expected that the population size in each cell is proportional to the capability $C_i$. Therefore, the populations are distributed in a heterogeneous environment.

In the first step of the dynamics, we randomly choose one site of the lattice to be colonized by a population that speaks the ancestor language. To each language, we assign a fitness value $f$ which is defined as the sum of the capabilities of the sites containing populations which speak that specific language. Therefore, the initial fitness of the ancestor language is the capability of the initial site. In the second step, one of the four nearest neighbors of this site will be chosen to be colonized with probability proportional to its capability. We assume that regions containing larger amount of resources are more likely to be colonized faster than poor regions. The referred site is then occupied by a population speaking the ancestor language or a mutant version of it. The assumption of mutations mimics the situation at which one language is initially spoken by populations in different regions, and after some time, modifications of this initial language emerge in one or both populations and the language split into two different languages. The probability of occurrence of a mutation in the process of propagation of the language is
$p=\frac{\alpha}{f}$, where $\alpha$ is a constant, and so the mutation probability is inversely proportional to the fitness of the language. This rule for the mutation procedure, we borrow from population genetics. We observe that small populations are more vulnerable to genetic drift and the rate of drift is inversely proportional to the population size \cite{Barton}. Genetic drift is a mechanism of evolution that acts to change the characteristics of species over time. It is a stochastic effect that arises due the random sampling in the reproduction process.

In the subsequent steps, we check what are the empty sites which are on the boundary of the colonized cluster, and we choose one of those empty sites according to their capabilities. Those ones with higher capabilities have a higher likelihood to be occupied. We then choose the language to occupy the cell among their neighboring sites. The languages with higher fitness have a higher chance to colonize the cell. This process will continue up to all sites be colonized. At this point, we verify the total number of languages $D$. In order to give to the reader some insight about our model, in Figure 1 we present the snapshot for a typical realization of the dynamics at the first moment of colonization of all sites. In this figure each color represent a different language domain. The color bar shows the label for each language.

\section{Results and Discussion}

\begin{figure}
\centering
\includegraphics[height=9cm,width=9cm,angle=0]{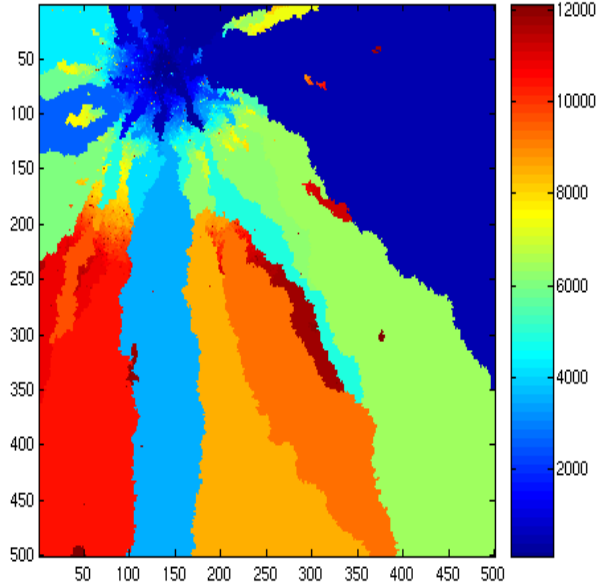}
\caption{Snapshot for a typical realization of the dynamics at the first moment of colonization of all sites. The size lattice was $L=500$ and $\alpha=0.73.$}
\end{figure}

In Figure 2, we show $D$ as a function of the area $A$ (total number of sites in the lattice) for two different values of the constant $\alpha$. We obtain each point by taking averages over 100 independent simulations for $L<300$, over 50 for $L=300$, 400 and 500 and over 20 for $L=700$. We notice from this figure the existence of two distinct scaling regions, where $D \sim A^z$. When $\alpha=0.3$, we estimate the exponent $z=0.43 \pm 0.04$ for $ 4 < A < 1,000$, whereas $z=0.14 \pm 0.02$ for $ 1,000 < A < 490,000$. When $\alpha=0.73$, we find $z=0.88 \pm 0.01$ for $4<A<10,000$, and $z=0.35 \pm 0.03$ when $10,000<A<490,000$. For both values of $\alpha$, we obtain exponents in agreement with those observed for the distribution of languages on Earth \cite{Gomes1}. Each power law extends over approximately two or more decades. For small and intermediate sized areas, the language diversity increases more quickly with the area, when compared to large areas. When $\alpha>0.9$, it is not possible to distinguish the two scaling regimes and in this case we estimate $z=1$. As the simulation is very time consuming for lattices of size $L>700$, it is not perfectly clear for us if the second regime is a true scaling or a transient regime with $D$ going to a constant value as $A$ increases or perhaps with $D$ growing logarithmically with $A$ for very large values of area.

\begin{figure}
\centering
\includegraphics[height=9cm,width=9cm,angle=270]{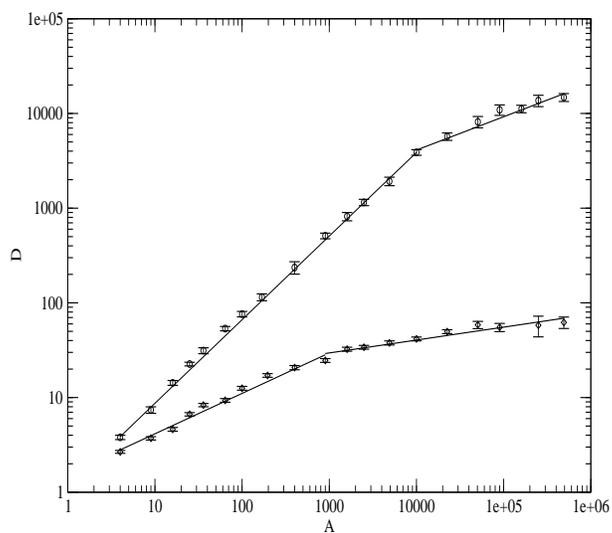}
\caption{ Number of languages $D$ as a function of the area $A$ for two values of mutation probability $p=\alpha/f$: $\alpha=0.3$, 0.73 (from bottom to top). The exponents obtained for $\alpha=0.3$ are $z=0.43 \pm 0.02$ for $4<A<1,000$, and $z=0.14 \pm 0.02$ for $1,000<A<490,000$. For $\alpha=0.73$ we have $z=0.88 \pm 0.01$ for $4<A<10,000$, and $z=0.35 \pm 0.03$ for $10,000<A<490,000$.}
\end{figure}

In order to characterize the diffusion in this process we investigate the time evolution of the average area $\bar{A}(t)$ occupied by the typical language. In our model, one time step represents the process of colonization of just one site by one language. The average area is

\begin{equation}
\bar{A}(t)= \frac{\sum_{i=1}^{D(t)}A_i(t)}{D(t)}
\end{equation}

\noindent
where $A_i(t)$ is the area occupied by the language $i$ in time $t$ and $D(t)$ is the diversity of languages in time $t$. By the end of the process of colonization we have $t=t_f=L^2$, and

\begin{equation}
\bar{A}(t_f) \sim L^{2(1-z)}
\end{equation}

\noindent
since the total area is equal to $L^2$ and the diversity is proportional to $A^{z}$. $\bar{A}(t_f)$ is a measure of the capacity of the languages to diffuse or to spread across the entire territory. Thus, the diffusion exponent $d_w$ can be introduced using the relation \cite{Havlin}

\begin{equation}
\bar{A}(t_f) \sim t_f^{2/d_w}.
\end{equation}

\noindent
From (2) and (3) we conclude that $d_w=2/(1-z)$, and thus $d_w$ assumes the standard Brownian value $d_w^B=2$ if $z=0$. For $0<z<1$ we have anomalous diffusion and $d_w>2$, indicating a progressive difficulty for language diffusion. For the current distribution of languages on Earth this exponent is $d_w= 3.4$, reflecting the geographical limitations that languages have to face in their process of expansion. This particular anomalous value is close to those obtained for diffusion in two-dimensional percolation below the percolation threshold \cite{Stauffer}. We compare our previous estimative (Eq. (3)) with the value of $d_w$ obtained by observing the evolution of $\bar{A}$ in the simulations and verify a good agreement between them. In Figure 3 we illustrate the evolution of $\bar{A}(t)$ for one run where $\alpha=0.3$ and $L=500$. For this particular situation we obtained $d_w = 2.08$, i. e. an exponent close to the Brownian value.

\begin{figure}
\centering
\includegraphics[height=9cm,width=9cm,angle=270]{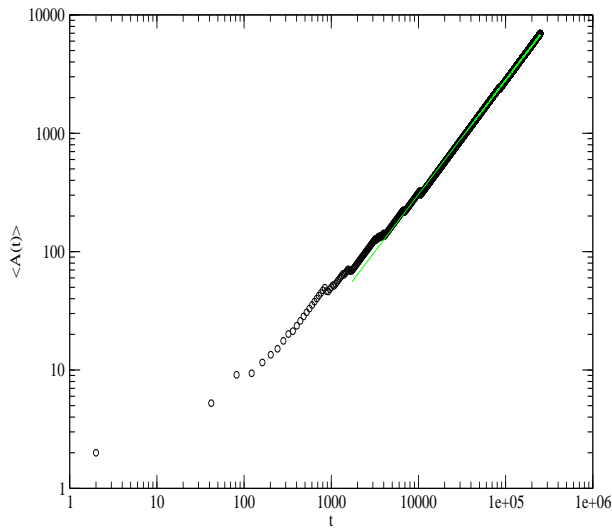}
\caption{ Evolution of $\bar{A}$ for $\alpha=0.3$ and $L=500$. For this curve, we have $d_w=2.08$.}
\end{figure}

\begin{figure}
\centering
\includegraphics[height=9cm,width=9cm,angle=270]{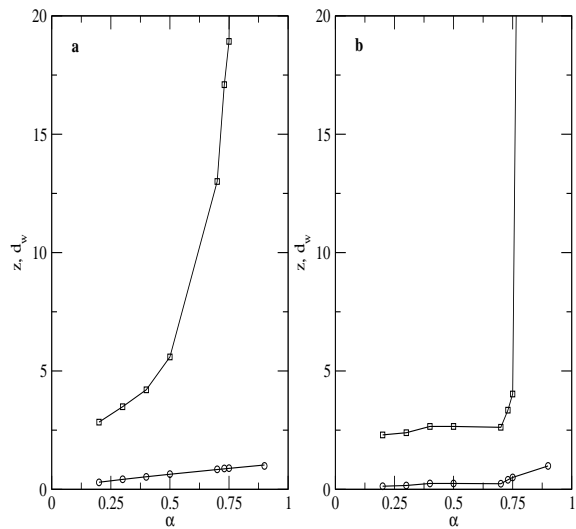}
\caption{ Exponents $z$ (o) and $d_w$ $(\square)$ as a function of $\alpha$ for (a) small and intermediate areas and (b) large areas.}
\end{figure}

\begin{figure}
\centering
\vspace{-0.5cm}
\includegraphics[height=9cm,width=9cm,angle=270]{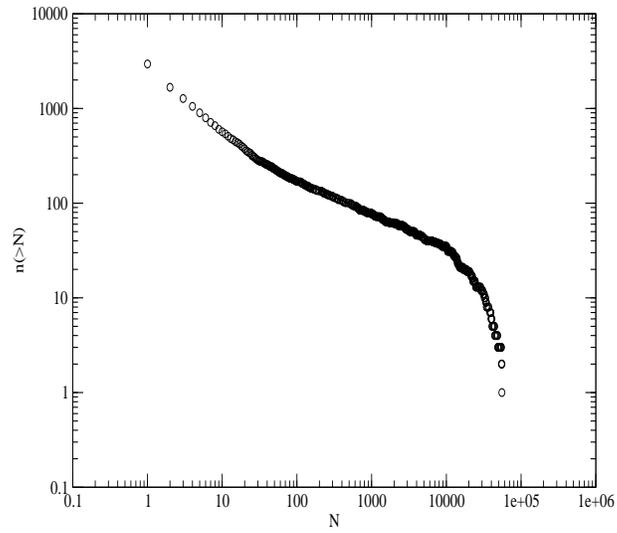}
\caption{\label{fig:figure3} Number of languages with population greater than $N$, $n(>N)$, as a function of $N$. $n(>N) \sim N^{\tau}$ with $\tau=0.36 \pm 0.01$ for $20<N<10,000$.}
\end{figure}

\begin{figure}
\centering
\includegraphics[width=9cm,height=9cm,angle=270]{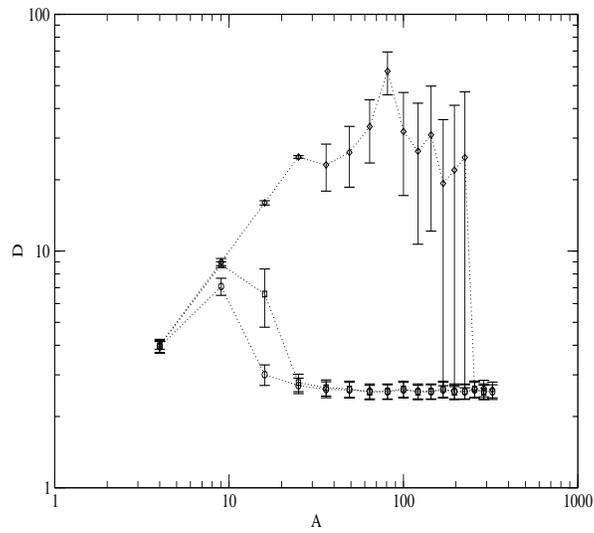}
\caption{\label{fig:figure4} Number of languages $D$ as a function of the area $A$ for $\alpha=0.73$ after the process of interaction among populations. The distinct curves show the cases where the fitness of the language that occupy the site is multiplied by 1, 10 and 100 (from bottom to top).}
\end{figure}

Figure 4 displays the dependence of the exponents $z$ and $d_w$ on the parameter $\alpha$: we observe that for small and intermediate areas, $z$ scales with $\alpha$ as $z \sim \alpha^{0.82 \pm 0.02}$, which means that a small increment of the mutation probability results in a fast increasing of the diversity. For large areas $z$ is approximately constant for $\alpha \le 0.7$ and quickly grows for $\alpha > 0.7$ (this sudden increase in $z$ possibly reflects the finite size of the lattice).

In Figure 5 we plot the number of languages with population size greater than $N$, $n(>N)$, as a function of $N$. In order to obtain the curves, we assume that each site contributes with 1 person to the population. The data points were estimated over 10 independent runs with $L=500$ and $\alpha=0.73$. In close analogy with the distribution of languages on Earth \cite{Gomes1}, we find the scaling regime $n(>N) \sim N^{-\tau}$ where $\tau=0.36 \pm 0.01$, along almost three decades in $N$.

\vspace{0.3cm}

After the populations have filled up the lattice, we initiate the second part of our analysis, which consider the process of interaction among populations. This interaction represents the flow of people that speak different languages among nations or even the people using new technologies which permit their communication throughout the world. Now, each time step corresponds to visit all cells on the lattice. In the visit to a given cell we compare the fitness of its language with the fitness values of languages which are spoken in its neighborhood. One of the five languages will be chosen to invade (or to stay) in that site with probability proportional to their fitness. The probability of mutation is, as before, proportional to the inverse of the fitness of the invading or staying language. After the stabilization, we estimate the average of the number of languages over a given time interval. In Figure 6 we plot the number of languages as a function of the area after considering the process of interaction. We obtain each point taking averages over 10 independent simulations. We also show the cases where the fitness of the language that is already in the site is increased by a factor 10 and 100 in order to compete with the neighbors. The purpose of this augment in the fitness is to increase the selective advantage of the population that already colonizes the site. When this selective advantage is small, the number of languages presents a linear growth and then decreases with the area up to reaching an asymptotic value. When the selective advantage is high, the number of languages initially presents a linear growth with the area, followed by a regime of fluctuation of the diversity and above a given area threshold it decreases abruptly.

For small areas, the fitnesses of populations are not high when compared to those one which can be obtained by populations in large areas. Therefore, in order to compete with the populations which already colonize the sites (and possess a selective advantage), languages need to colonize large areas. This is the reason why we observe the initial linear growth of language diversity for small area when we consider a high selective advantage. Above a certain area, some populations have a very high fitness in order to compete with other languages and dominate. Thus, the diversity drastically decreases and only a few languages survive.

\section{Conclusions}

We have introduced in this work a simple computer model to simulate some aspects of the linguistic diversity on Earth. Surprisingly, this model is able to generate important scaling laws (Figs. 2 and 5) in close resemblance to those observed in the actual distribution of languages \cite{Gomes1}.

We verified that the mutation probability of languages displays a decisive role for the maintenance of the linguistic diversity. On the other hand, the diffusion exponent $d_w$ is very large for values of $\alpha$ close to 1 (Fig. 4), that is, in the situation in which we have a high diversity per area, indicating that languages do not have much facility to diffuse and remain essentially localized in linguistic niches.

We do not discard the possibility that the presence of two scaling regimes for the linguistic diversity in our simulations may be a consequence of the regular structure of the lattice. In order to investigate the role of the topology in this model, we are currently studying the process of diffusion of languages in a percolation cluster.

After the process of interaction among languages we observed the surviving of just a few languages with a very high fitness (Fig. 6). From this, we can ascertain how important is for linguistic diversity to have well stablished languages spoken by a large number of people.

V.~M. de Oliveira and M.~A.~F. Gomes are supported by Conselho Nacional de Desenvolvimento Cient\'{\i}fico e Tecnol\'ogico and Programa de N\'ucleos de Excel\^encia (Brazilian Agencies).


\begin{thebibliography}{99}

\bibitem{Renfrew2}{C. Renfrew, Man {\bf 27}, No. 3, 445 (1992).}

\bibitem{Renfrew}{C. Renfrew, Scientific American {\bf 270}, 104 (1994).}

\bibitem{Cavalli-Sforza2}{Cavalli-Sforza, L. L. {\it Genes, Peoples and Languages} (Penguin, London, 2001).}

\bibitem{Renfrew3}{C. Renfrew, P. Forster, M Hurles, Nature Genetics {\bf 26}, 253 (2000).}

\bibitem{Cavalli-Sforza}{L. L. Cavalli-Sforza, Scientific American {\bf 265}, 72 (1991).}

\bibitem{Diamond}{J. M. Diamond, Nature {\bf 389}, 544 (1997).}

\bibitem{Sampat}{P. Sampat, World Watch (may/june 2001), 34.}

\bibitem{Graddol}{D. Graddol, Science {\bf 303}, 1329 (2004).}

\bibitem{Gomes1}{M. A. F. Gomes, G. L. Vasconcelos, I. J. Tsang, I. R. Tsang, Physica A {\bf 271}, 489 (1999).}

\bibitem{Rosenzweig}{M. L. Rosenzweig, Species Diversity in Space and Time (Cambridge University Press, Cambridge, 1995).}

\bibitem{Pelletier}{J. D. Pelletier, Phys. Rev. Lett. {\bf 82}, 1983 (1999).}

\bibitem{Hanski}{I. Hanski, M. Gyllenberg, Science {\bf 275}, 397 (1997).}

\bibitem{Barton}{N. H. Barton, in Speciation and Its Consequences, edited by D. Otte and J. A. Endler (Sinauer Associates, Inc., Sunderland, MA, 1989), pp 229-256.}

\bibitem{Havlin}{S. Havlin, D. Ben-Avraham, Adv. Phys. {\bf 51}, 187 (2002).}

\bibitem{Stauffer}{Stauffer, D., Introduction to Percolation Theory (Taylor $\&$ Francis, London, 1985).}




\end{thebibliography}
\end{document}